\documentclass[sigconf]{acmart}
\usepackage[]{collab}
\usepackage{enumitem}
\usepackage{balance}
\usepackage{xcolor}
\usepackage{subfigure}
\usepackage{multirow}
\usepackage[most]{tcolorbox}
\usepackage{tabularx}

\newcommand{\etal}{{\em et al.}\xspace}
\newcommand{\ie}{{\em i.e.},\xspace}
\newcommand{\eg}{{\em e.g.},\xspace}

\definecolor{ballblue}{rgb}{0.13, 0.67, 0.8}

\collabAuthor{jy}{ballblue}{Jinyang Liu}
\collabAuthor{jz}{magenta}{Jiazhen Gu}
\collabAuthor{jj}{purple}{Junjie}
\collabAuthor{jx}{blue}{Jinxi}
\collabAuthor{ry}{pink}{Renyi}

\newcommand\frnm{COLA\xspace}
\newcommand\stnm{correlation mining module\xspace}
\newcommand\smnm{LLM reasoning module\xspace}

\newcommand\huawei{Cloud X\xspace}

\copyrightyear{2024}
\acmYear{2024}
\setcopyright{acmlicensed}\acmConference[ICSE-SEIP '24]{46th International Conference on Software Engineering: Software Engineering in Practice}{April 14--20, 2024}{Lisbon, Portugal}
\acmBooktitle{46th International Conference on Software Engineering: Software Engineering in Practice (ICSE-SEIP '24), April 14--20, 2024, Lisbon, Portugal}
\acmDOI{10.1145/3639477.3639745}
\acmISBN{979-8-4007-0501-4/24/04}

\AtBeginDocument{%
  \providecommand\BibTeX{{%
    \normalfont B\kern-0.5em{\scshape i\kern-0.25em b}\kern-0.8em\TeX}}}

\begin{document}

\title{Knowledge-aware Alert Aggregation in Large-scale \\ Cloud Systems: a Hybrid Approach}

\author{Jinxi Kuang$^\S$, Jinyang Liu$^\S$, Junjie Huang$^\S$, Renyi Zhong$^\S$,\\ Jiazhen Gu$^\S{}^*$, Lan Yu$^\|$, Rui Tan$^\|$, Zengyin Yang$^\|$,  Michael R. Lyu$^\S$}
\affiliation{%
  \institution{$^\S$The Chinese University of Hong Kong, Hong Kong SAR, China,\\ \{jxkuang22, jyliu, jjhuang23, ryzhong22, jiazhengu, lyu\}@cse.cuhk.edu.hk}
  \country{}}
\affiliation{%
  \institution{$^\|$Computing and Networking Innovation Lab, Huawei Cloud Computing Technology Co., Ltd, China,\\ \{yulan13, tanrui3, yangzengyin\}@huawei.com}
  \country{}}

\renewcommand{\shortauthors}{Jinxi Kuang, et al.}

\begin{abstract}

Due to the scale and complexity of cloud systems, a system failure would trigger an "alert storm", \ie massive correlated alerts. Although these alerts can be traced back to a few root causes, the overwhelming number makes it infeasible for manual handling.
Alert aggregation is thus critical to help engineers concentrate on the root cause and facilitate failure resolution.
Existing methods typically utilize semantic similarity-based methods or statistical methods to aggregate alerts. However, semantic similarity-based methods overlook the causal rationale of alerts, while statistical methods can hardly handle infrequent alerts.

To tackle these limitations, we introduce leveraging external knowledge, \ie Standard Operation Procedure (SOP) of alerts as a supplement.
We propose \frnm, a novel hybrid approach based on correlation mining and LLM (Large Language Model) reasoning for online alert aggregation.
The correlation mining module effectively captures the temporal and spatial relations between alerts, measuring their correlations in an efficient manner.
Subsequently, only uncertain pairs with low confidence are forwarded to the LLM reasoning module for detailed analysis.
This hybrid design harnesses both statistical evidence for frequent alerts and the reasoning capabilities of computationally intensive LLMs, ensuring the overall efficiency of \frnm in handling large volumes of alerts in practical scenarios.
We evaluate \frnm on three datasets collected from the production environment of a large-scale cloud platform.
The experimental results show \frnm achieves F1-scores from 0.901 to 0.930, outperforming state-of-the-art methods and achieving comparable efficiency.
We also share our experience in deploying \frnm in our real-world cloud system, \huawei \footnote{Due to the company policy, we anonymize the name as \huawei.}.
\renewcommand{\thefootnote}{}
\footnotetext{*Corresponding author.}

\end{abstract}



\maketitle

\section{Introduction}

Cloud systems, such as Microsoft Azure, Amazon AWS, and Google Cloud Platform (GCP), have been providing essential services for customers worldwide. 
It is crucial to ensure the reliability of these systems in order to prevent user dissatisfaction and economic loss. For example, the cost of downtime for Amazon in an hour on Prime Day is up to 100 million dollars~\cite{chen2019empirical}.

\begin{figure}[t]
  \centering
  \includegraphics[width=0.44\textwidth]{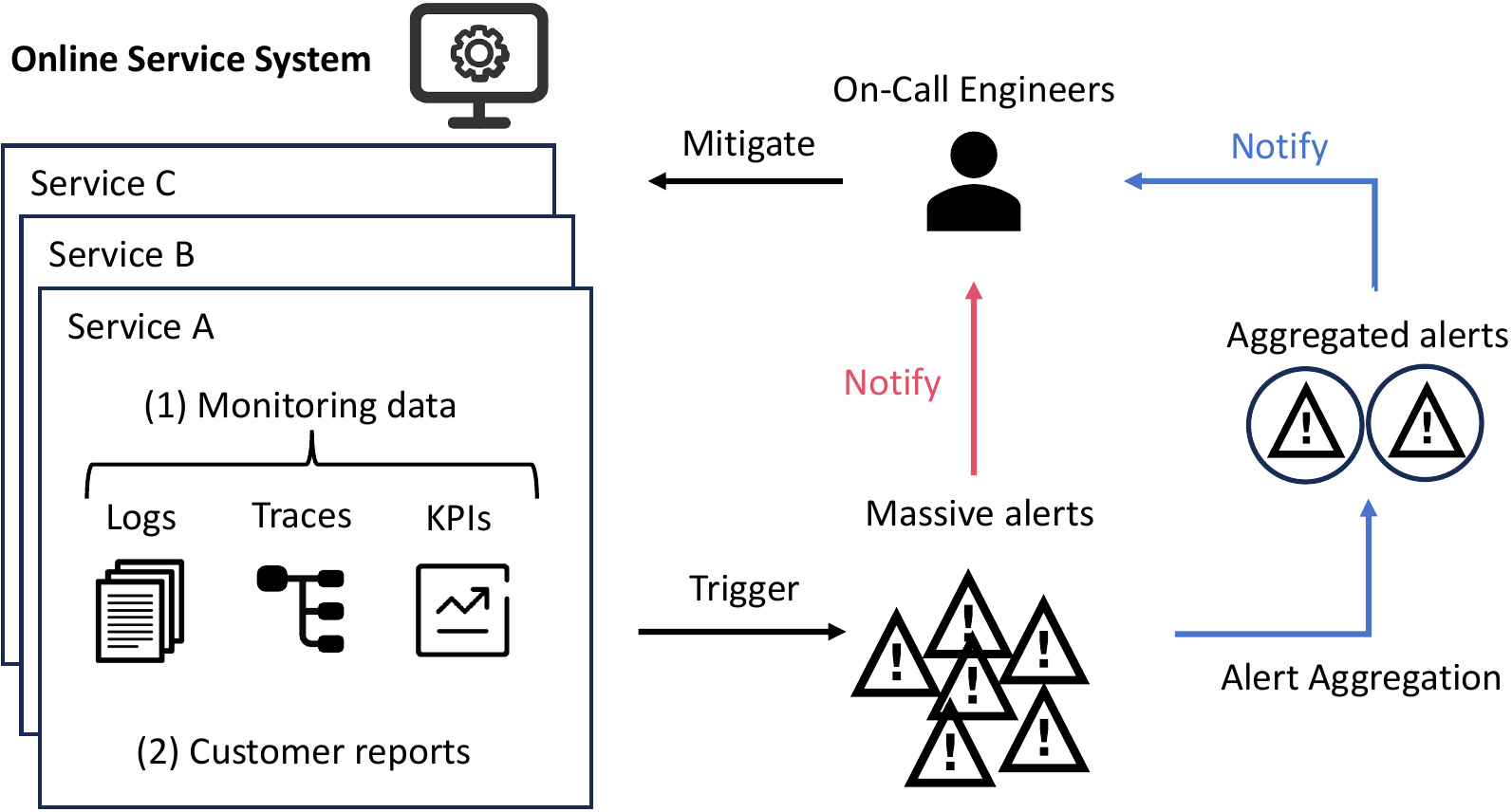}
  \caption{The alert handling process}
  \label{fig:bg}
  \vspace{-10pt}
\end{figure}

In practice, failures (\eg unexcepted interruptions or service level degradation) are inevitable in cloud systems~\cite{chen2019empirical, chen2020incidental, jiang2020mitigate}.
To detect and handle failures promptly, modern cloud systems have configured comprehensive monitoring mechanisms to monitor the health status of cloud services continuously, which produces monitoring data including KPI (key performance indicators), logs and traces~\cite{gu2023performance, peng2022revisiting, liu2019logzip, liu2023scalable, lee2023eadro, lee2023heterogeneous}. 
When an unexpected pattern is detected over these data, an \textit{alert} will be triggered to notify the On-Call Engineers (OCEs) for inspection and mitigation as demonstrated by the red path in Figure~\ref{fig:bg}.
For example, when the number of failed requests increases over a threshold, an alert ``failed request number exceeded 100 within 60 seconds'' would be fired to an OCE. 

When a failure happens, massive correlated alerts will be triggered in a brief time period, a phenomenon typically referred to as the notorious "alert storm" problem~\cite{zhao2020alertstorm}.
This issue primarily arises from the large-scale and complex dependency of modern cloud systems. 
Specifically, a cloud system can own and monitor tens of hundreds of services~\cite{chen2020towards, chen2021graph}. Furthermore, these services are interdependent, meaning a failure in one service has the potential to spread to others, a process known as cascading failures~\cite{he2018identifying}.
For instance, a failure that occurs within a database service could potentially impact other services that depend on it such as user authentication, transaction processing, or data analytics.
In this process, alerts are usually triggered in an uncoordinated manner, which are overwhelming for OCEs for manual handling.
To address this issue, it is crucial to \textit{ automatically aggregate these alerts that are caused by the same failure (\ie having the same root cause)}, which can improve the efficiency of OCEs in resolving the failure as depicted by the blue path in Figure~\ref{fig:bg}.

Many efforts have been devoted to alert aggregation, which can be categorized into semantic similarity-based methods and statistical methods.
Semantic similarity-based methods, such as AlertStorm~\cite{zhao2020alertstorm}, LiDAR~\cite{chen2020lidar} and OAS~\cite{chen2022oas}, correlate two alerts having similar semantics based on nature language processing technologies (\eg word2vec and BERT). 
However, correlated alerts can have distinct semantics. For example, an alert warning of a server overload may be correlated with an alert indicating a database slowdown, even though the semantics of these two alerts are quite different. 
On the other hand, statistical methods, such as Warden~\cite{li2021fighting}, LiDAR~\cite{chen2020lidar} and iPACK~\cite{liu2023ipack}, learn from the co-occurrence patterns of two alerts to determine their relations.
Nevertheless, such methods rely on a large amount of historical data for learning, which might be inaccurate for unseen or infrequent alerts.

To address the limitations of existing studies, motivated by the recent successes of large language models (LLM) for challenging reasoning tasks~\cite{ahmed2023recommending, chen2023empowering, jin2023assess}, we propose to leverage LLM for alert aggregation in a large-scale cloud system. 
Intuitively, we aim to perform reasoning about relations between two alerts using LLM by learning from existing knowledge of alerts.
To be specific, in \huawei, each alert is associated with a comprehensive document called Standard Operating Procedure (SOP).
These SOPs contain detailed information about the alert, such as the conditions that trigger it, severity levels, potential impacts, possible root causes, and recommended mitigation steps.
The SOPs are continually updated by on-site engineers during maintenance activities, \eg when they observe a new potential impact.
These SOPs provide insights into the underlying rationale behind alerts, enabling us to engage in reasoning rather than relying solely on semantic similarity comparisons.
Furthermore, LLMs can generate interpretable results for on-site engineers, facilitating their understanding the outcomes.

However, designing such a solution faces the following three challenges.
First, existing LLMs are not specialized for alert aggregation. 
These models are typically trained using publicly available general knowledge sources, such as Wikipedia and Github.
Consequently, they lack specific knowledge about alerts in cloud systems, which is essential for accurately reasoning about the relationships between alerts.
Second, the SOPs typically consist of extensive textual content. Considering that LLMs often struggle to effectively learn from lengthy input~\cite{zhao2023survey}, performing reasoning directly on such long text poses a significant challenge for these models.
Third, the efficiency of LLMs is low. It generally takes seconds for a typical LLM (\eg ChatGPT) to generate responses. In real-world deployment scenarios, this time delay is far from ideal, particularly when a significant number of alerts are generated during a failure event.

\textbf{Our work.} To tackle these challenges, in this paper, we propose \textbf{\frnm}, a novel hybrid approach based on \uline{CO}rrelation mining and \uline{L}LM reasoning for online alert \uline{A}ggregation. 
\frnm is composed of two modules, a correlation mining module and an LLM reasoning module.
To tackle the first challenge, we propose to leverage in-context learning (ICL) to provide domain-specific knowledge to LLM. ICL is prompt engineering that demonstrates relevant examples to LLM to enable LLM to perform tasks by learning from the examples, which has been proven to be effective in various software engineer tasks~\cite{zheng2023towards,hou2023large}.
To address the second challenge, we propose a multi-round prompting mechanism to summarize long text within the SOPs progressively and extract key knowledge for reasoning relations between alerts.
For the third challenge, we propose to incorporate a lightweight and efficient correlation mining component in our framework. This component aims to filter out correlated alert pairs with high confidence. Only those uncertain pairs are sent to the LLM reasoning module for in-depth analysis.
To evaluate the proposed approach, we collect three datasets from the real-world production environment of \huawei, which contain 500,000 alerts and 3,000 SOPs. Extensive experiments demonstrate that \frnm achieves F1-scores from 0.901 to 0.930, significantly outperforming all state-of-the-art baseline models. Furthermore, despite utilizing computationally intensive LLM for analysis, \frnm demonstrates comparable efficiency to existing methods.

Our major contributions are as follows:
\begin{itemize}[leftmargin=*]
    \item We are the first to propose introducing detailed knowledge (\eg SOPs) for alert aggregation. By leveraging this extensive knowledge, we are able to effectively aggregate alerts that are semantically dissimilar or infrequent (Section~\ref{sec:background}).
    \item We propose \frnm, a hybrid framework comprising a correlation mining component and an LLM reasoning component, specifically designed for online alert aggregation. This hybrid approach allows us to harness the capabilities of LLM while ensuring efficient handling of a large volume of alerts in practical scenarios (Section~\ref{sec:method}).
    \item We evaluate \frnm on three datasets collected from the industrial production environment. The evaluation results show that \frnm outperforms the state-of-the-art method with F1-scores from 0.901 to 0.930 and archives comparable efficiency (Section~\ref{sec:evaluation}). 
    \item We have deployed \frnm in the production environment of \huawei for four months. Our practical experience is shared to benefit the community (Section~\ref{sec:indst}).
\end{itemize}

\section{Background and Motivation}
\label{sec:background}

In this section, we introduce background concepts in \huawei, and demonstrate a motivating example to elaborate our motivation.

\subsection{Alerts and SOPs in \huawei}

\begin{figure}[t]
  \centering
  \includegraphics[width=0.45\textwidth]{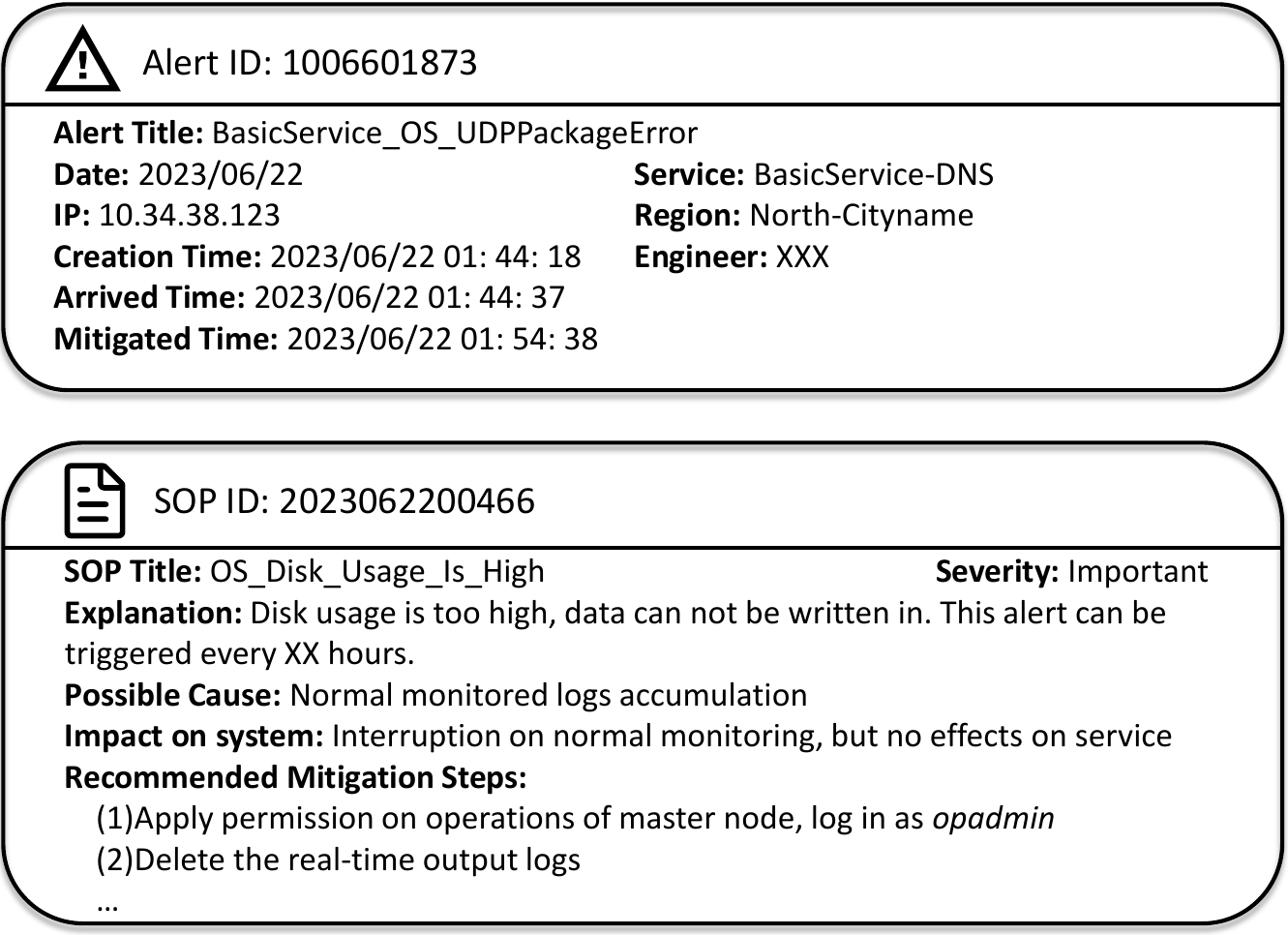}
  \caption{Example of an alert and a SOP document}
  \vspace{-20pt}
  \label{fig:alertsop}
\end{figure}

\noindent{\textbf{Alert.}}
Alerts are produced when anomalies are detected in the monitoring data or raised by customers~\cite{zhao2020alertstorm}. 
Alerts are used to notify OCEs for timely anomaly handling and have various attributes including alert ID, title, creation time, arrival time, mitigated time, owning service, region and corresponding engineer as Figure~\ref{fig:alertsop} shows. 
\textit{ID} is unique for each alert, and the \textit{title} can be regarded as a summary of the alert. 
\textit{Creation time}, \textit{arrival time}, \textit{mitigated time} denote the specific time when an alert is triggered, received by OCEs, and mitigated by OCEs, respectively.
\textit{Service} shows the owning service of alert, and \textit{region} indicates the physical location of the server where the service is deployed.
\textit{Engineer} contains the name and ID of the responsible OCE.

\begin{table*}[]
\small
\captionsetup{justification=centering}
\centering
\caption{Correlated alerts and associated SOPs in \huawei: a motivating example}
\vspace{-5pt}
\label{tab:moti}
\begin{tabular}{|c|c|c|c|}
\hline
\multicolumn{2}{|c}{\textbf{Alert}} & \multicolumn{2}{|c|}{\textbf{SOP (after manual summarization)}}\\
\hline
\textbf{ID} & \textbf{Title} & \textbf{Impact on the system} & \textbf{Possible causes}  \\ 
\hline
$a_{1}$ & SSD voltage is lower than alarming threshold. & Latency of memory processes increases & SSD failure \\
\hline
$a_{2}$ & OSD I/O blocking detected. & Latency of disk requests increases & Hardware failure or OS bugs \\
\hline
$a_{3}$ & OSD process exception detected.& Data reconstruction on OSD managed disks & Unreadable sectors on disk \\
\hline
$a_{4}$ & OSD node 19e7fb37 is unhealthy. & Latency of disk requests increases or fluctuates & OSD I/O not responding \\
\hline
$a_{5}$ & Storage cluster a9dbc69b is abnormal. & Storage pool redundancy reduction or failure & OSD nodes are abnormal  \\
\hline
$a_{6}$ & Successful connection count to Redis is decreased. & Disconnection to Redis service & Hardware failure or component abnormal\\
\hline
\end{tabular}
\end{table*}

\noindent{\textbf{SOP.}}
The Standard Operating Procedure (SOP)~\cite{yang2022sop} documents are alert description documents maintained by engineers in \huawei. 
The documents are written in natural language along with screenshots, tablets and website links to provide a reference for alert handling. 
SOPs contain the ID, title, severity, explanation of the alert, the impact on the system, possible cause, and recommended mitigation steps as shown in Figure~\ref{fig:alertsop}.
\textit{ID} is the identifier and \textit{title} briefly introduces the anomaly.
\textit{Severity} indicates the priority of handling.
\textit{Explanation} elaborates more detail about this alert.
\textit{Impact} shows the consequential results of this alert on the owning service and the whole system.
\textit{Possible cause} and \textit{recommended mitigation steps} are concluded by engineers from similar historical alerts.
In addition, the domain knowledge of cloud services is also implicitly contained in the engineer-write notes and explanations. 
However, SOPs are long and unstructured.
Thus it is challenging to extract the key information automatically and further analyze the correlation.

\subsection{Motivating Example}

We present a real-world service failure in January 2023 with part of the relevant alerts and SOP in \huawei as a motivating example shown in Table~\ref{tab:moti}.
The failure was originally caused by the low voltage on the solid-state drive (SSD in a storage server ($a_1$), leading to the interruption of Object Storage Device (OSD) processes. 
Specifically, some of the I/O operations of OSD were incomplete or blocked ($a_2$) due to insufficient power supply in SSD.
This was captured by the OSD process monitors as an exception ($a_3$), which resulted in an unhealthy OSD node ($a_4$).
Next, the health monitor for the storage cluster also reported such anomalies ($a_5$).
Finally, at the application layer, we observed successful connection to our Redis service is decreased ($a_6$).
In practice, a lot more alerts can be generated in a large-scale cloud system; we only selected representative ones for clearer demonstration.

Correlating these overwhelming alerts poses a significant challenge for OCEs, as it demands extensive experience and understanding of cloud systems to interpret alerts originating from various services across different levels of the infrastructure. 
In this example, alerts $a_1 \sim a_6$ cover hardware at the infrastructure level, nodes and processes at the platform level, and the application level.
Previous studies rely on statistical correlation or semantic similarity to perform alert aggregation, which, however are suboptimal.
For alerts shown in Table~\ref{tab:moti}, not all of them appear extensively to support statistical analysis. For example, $a_5$ and $a_6$ are rare since they generally represent more severe problems. In addition, $a_{1}\sim a_{6}$ share little semantic similarities.

We propose to perform reasoning based on knowledge (\eg SOPs) of each alert for more effective alert aggregation. 
SOPs associated with each alert in the cloud system are detailed and comprehensive, spanning approximately 3 to 4 A4 pages. 
For clarity, we have provided summarized versions of these SOPs in Table~\ref{tab:moti}.
Inspired by the recent success of LLM for various challenging tasks~\cite{huang2022towards}, we aim to utilize LLM to comprehend these SOPs, extract key information and aggregate correlated alerts based on reasoning. 
In this way, we can link alerts based on extensive knowledge and obtain interpretable results from LLM. 
Notably, this approach could leverage the knowledge distributed across different services and layers of the cloud system, and perform more effective alert aggregation.

\subsection{Challenges}
Although LLM is capable of extracting knowledge from natural language and performing further reasoning tasks, there are still three challenges of LLM that should be addressed in practice.

\noindent\textbf{Challenge 1: Low efficiency.}
LLM is time-consuming to generate a response in practical industrial scenarios. 
It takes LLM more than 10 seconds to summarize an SOP and even more time to perform reasoning in \huawei.
When service failure happens, a large number of alerts are triggered, making it impossible to handle the failure timely by inputting all the alerts to LLM. 
To overcome the low-efficiency challenge, we first filter the alerts with the similarity score evaluated by the correlation mining module. 
Then We drop out the alert pairs with positive similarity scores, and the LLM module only needs to process a small scale of undetermined alerts.

\noindent\textbf{Challenge 2: Long text understanding.}
The context understanding and summarizing ability of LLM decreases as the length of the text grows.
Most of the input lengths of LLM are limited. 
Some LLMs claim they support infinite input length, but they also stress the effectiveness would decline after inputting thousands of words. 
In our design, the prompt for LLM contains two alerts and SOPs, which is usually longer than these limits. 
To overcome the long-text understanding challenge, inspired by the Chain-Of-Thought (COT) design~\cite{wei2022chain}, we propose a two-round interaction with LLM. In the first round of requests, we ask LLM to extract and summarize the knowledge in SOP; In the second round, we ask LLM to perform analyzing and reasoning on alerts with the summarized SOP.

\noindent\textbf{Challenge 3: Lack of domain knowledge.}
LLM is a generalized language model that lacks of domain knowledge of online service systems.
LLM is trained using a huge amount of accessible data on the Internet.
However, the knowledge of online service system maintenance, especially for those internal terms and descriptions, is mostly not publicly available. 
Thus it is difficult for LLM to generate a reasonable answer for causal analysis of alerts. 
To overcome the challenge of lacking domain knowledge, we apply two techniques to leverage the summarized knowledge from SOP. 
In-Context Learning (ICL) performs a few-shot learning. 
A few positive and negative examples, which contain summarized SOP and expected answers, are inserted into the prompt. LLMs are expected to quickly pick up the domain knowledge from those examples. 
Supervised Fine-Tuning (SFT) takes historical alerts and SOPs as the training set and learns the domain knowledge by updating the model parameters.

\section{Methodology}
\label{sec:method}

\begin{figure}[t]
  \centering
  \includegraphics[width=0.45\textwidth]{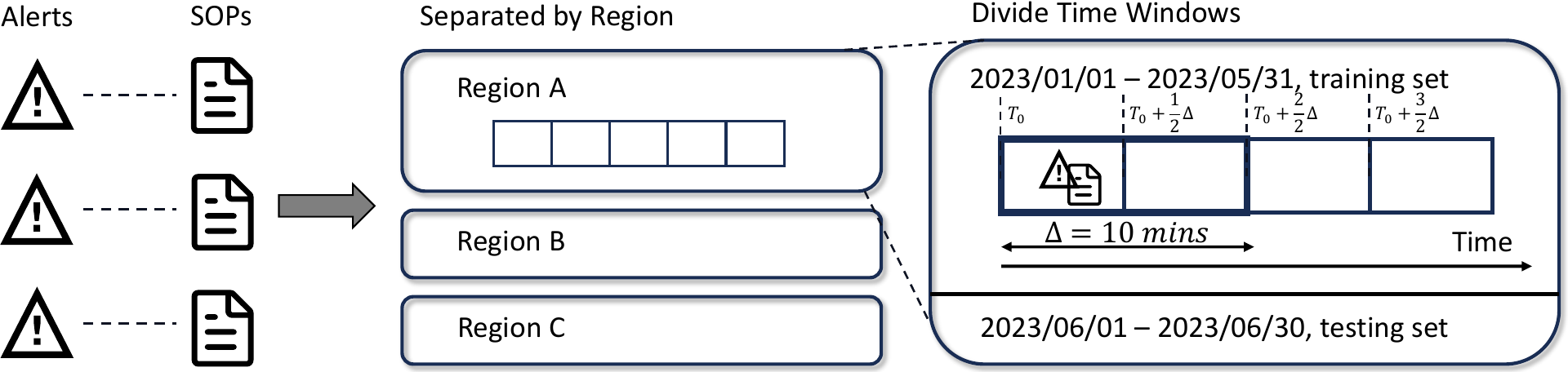}
  \caption{The preprocessing steps}
  \vspace{-15pt}
  \label{fig:prepro}
\end{figure}

\begin{figure*}[t]
  \centering
  \includegraphics[width=0.9\textwidth]{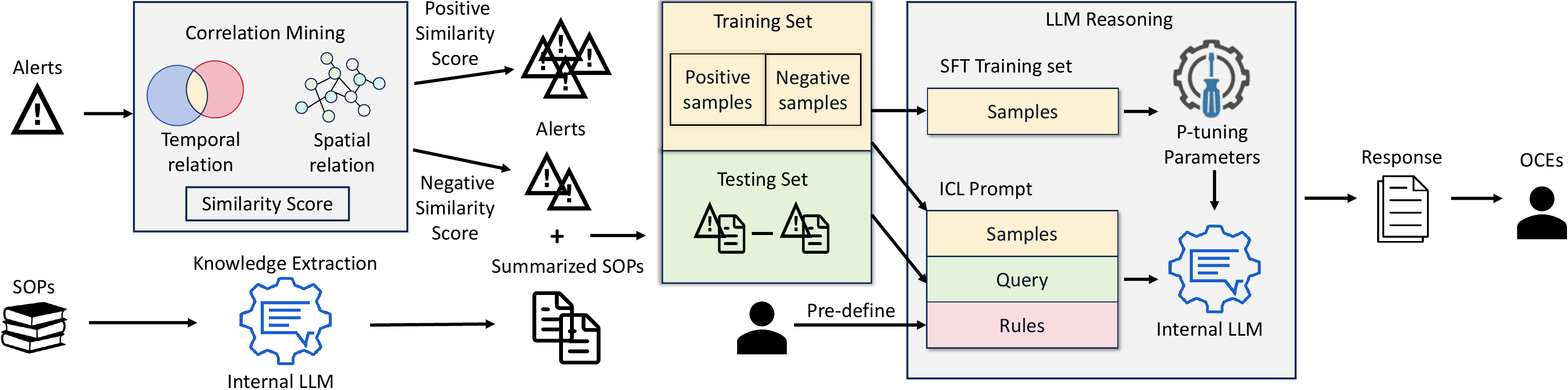}
\vspace{-10pt}
  \caption{Overview of \frnm}
  \vspace{-10pt}
  \label{fig:fr}
\end{figure*}

\subsection{Overview of \frnm}

The goal of \frnm is to aggregate the alerts online by performing correlation mining on statistic information and LLM reasoning on semantic information. The inputs of \frnm are alerts and the corresponding SOPs. The results are the grouped alerts, where the alerts have the same cause that each of the two alerts are correlated. The aggregation helps to reduce the number of alerts to handle, decreasing the workload of OCEs. In addition, compared to one single alert, the aggregated alerts are more beneficial to the OCEs in understanding the whole failure process. This also reduces the time cost on locating and fixing the service failure.

Figure~\ref{fig:fr} shows the overall framework of \frnm. \frnm first preprocesses the alerts obtained from the real-world scenarios. Those alerts are separated based on different physical regions which can be regarded as isolated systems, then are divided into time windows according to the \textit{Arrived Time} attribute. Next, the \stnm is proposed to determine the correlation of alert pairs by utilizing the temporal and spatial relation, and output a similarity score of correlation analysis. The alert pairs with a positive similarity score would be marked as correlated, while the alert pairs with negative scores would be processed in the \smnm along with the corresponding summarized SOPs. In \smnm, two popular LLM techniques, In-Context Learning (ICL) and Supervised Fine-Tuning (SFT), are leveraged to learn domain knowledge and perform reasoning tasks. In the ICL method, similar historical alerts and SOP samples are captured by FastText~\cite{joulin2016bag} embedding and semantic similarity search. Those samples as well as pre-defined reasoning rules are inserted into prompts of the current query to provide domain knowledge and inference examples for LLM. In the SFT method, the historical data is used for LLM fine-tuning, which makes LLM capable of performing alert-related reasoning with domain knowledge.

\subsection{Preprocessing}

In a large-scale system, services are deployed in various physical regions, which are isolated and can be regarded as independent. Thus we separate the alerts by regions at the first step as Figure~\ref{fig:prepro} shows, which reduces the workload for correlation mining and eliminates the error of grouping alerts from irrelevant services.

In most cases, the alerts of the same root cause are triggered in a certain period~\cite{zhao2020alertstorm}. This indicates the aggregation can focus on a relatively short time interval (e.g. several minutes). We call these intervals \textit{time windows}. According to our practical experience in \huawei, the length of time windows can be set to 10 minutes. Considering service failure can happen near the boundary of the interval, we employ the sliding window policy. The sliding size is set as $\frac{1}{2}$ windows. 
Formally, denote $T_{0}$ as the start time, $\Delta$ as the length of the time window, $s$ as the sliding size, then the time window for our alert division is: $[T_{0}+ks\Delta,\ T_{0}+(ks+1)\Delta], k \in [1,2,3...]$
With this division, we maintain the potential alert groups and also reduce the input scale for correlation mining. 
Finally, we split the training set and testing set based on the time attribute of alerts.
We further demonstrate this part in Section~\ref{sec:evaluation}.

\subsection{Correlation Mining Module}

In the correlation mining module, our objective is to identify alerts that exhibit significant statistical evidence of correlation efficiently.
In this way, when an alert storm happens, we can quickly filter out alert pairs with high confidence and only leave uncertain pairs for detailed analysis based on LLM.

To achieve this, the proposed correlation mining module incorporates both \textit{temporal relation} and \textit{spatial relation} for alert analysis.
For temporal relation, we consider the correlation in time windows. It is intuitive that the frequently co-occurring alerts are more likely to be correlated. If two alerts have the same root cause, they would be triggered in a short period as the alert storm occurs. Thus it can be captured and observed by the sliding windows. 
For spatial relations, we consider the propagation sequences within the service topology. If two services are not related in topology, the alerts from these two services should be more likely irrelevant, and vice versa. Combining the temporal and spatial relation, we obtain the statistically correlated alert pairs in this step.

\subsubsection{Leveraging Temporal Relations}

\begin{figure}[t]
  \centering
  \includegraphics[width=0.45\textwidth]{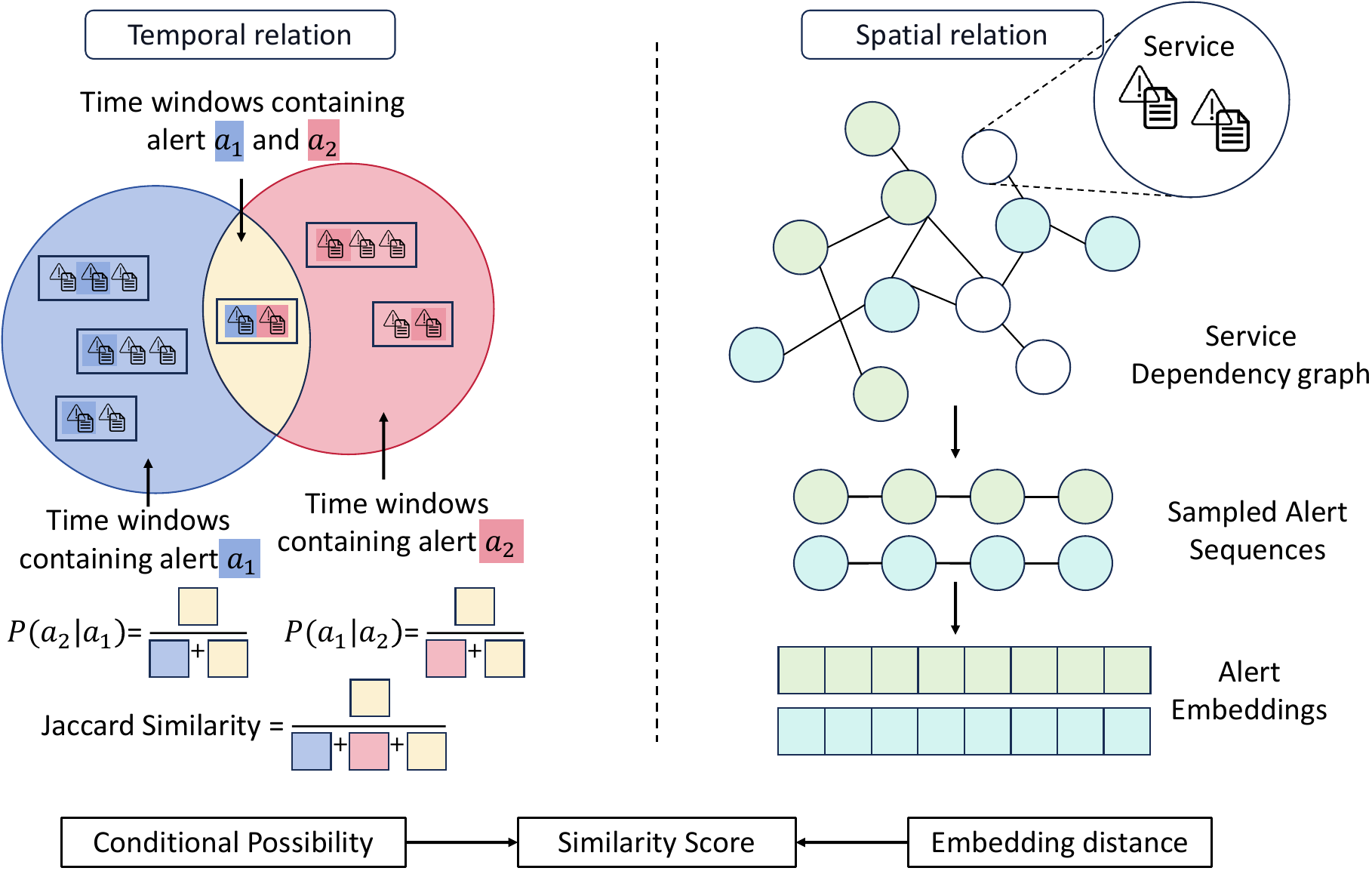}
  \caption{The design of correlation mining component}
  \label{fig:corr}
  \vspace{-20pt}
  
\end{figure}

To evaluate the correlation of alerts in time windows, we utilize conditional possibility as our metrics for frequent pattern mining. For the alert pair which contains alerts $a_{1}$ and $a_{2}$, we first collect all the time windows where alert $a_{1}$ is involved, to compute its occurrence frequency $P(a_{1}) = \frac{\#Windows\ containing\ a_{1}}{\#All\ windows}$. Similarly, we obtain $P(a_{2})$. Then we find the windows where $a_{1}$ and $a_{2}$ appear together to get $P(a_{1}a_{2}) = \frac{\#Windows\ containing\ a_{1}\ and\ a_{2}}{\#All\ windows}$. With these metrics, we can calculate the occurrence frequency of $a_{2}$ given that $a_{1}$ has occurred by $P(a_{2}|a_{1}) = \frac{P(a_{1}a_{2})}{P(a_{1})}$ and $P(a_{1}|a_{2}) = \frac{P(a_{1}a_{2})}{P(a_{2})}$ as the left part of Figure~\ref{fig:corr} illustrates. Thus we have two conditional possibilities that indicate the similarity from the temporal view denoted as $T_{a_{1}|a_{2}}, T_{a_{2}|a_{1}}$.

However, the alerts are not ensured to be correlated even if the conditional possibility $T_{a_{1}|a_{2}}$ or $T_{a_{2}|a_{1}}$ is high. Some alerts are issued regularly to act as a reminder during the normal execution of systems. These regular alerts are regarded as noises in our task, which cannot represent a practical co-occurrence pattern in aggregation, and those alerts would further affect the performance of correlation analysis. Thus it is significant to remove the noise alerts in our collected dataset.   

To detect the noise in alerts, we utilize the Jaccard similarity to evaluate the proportion of overlapping for the two sets of time windows. The Jaccard similarity is calculated as dividing the cardinality of the intersection by the cardinality of the union. With the notations in this section, it can be denoted as $Jaccard\ similarity\ = \frac{P(a_{1}a_{2})}{P(a_{1}+a_{2})} = \frac{P(a_{1}a_{2})}{P(a_{1}) + P(a_{2}) - P(a_{1}a_{2})}$. The Noise alert $a_{N}$ appear in time windows evenly, leading to $P(a_{N}a_{i})\approx P(a_{i})$ and $P(a_{N}a_{i}) \ll P(a_{N})$, for most of the non-noise alert $a_{i}$. Thus the Jaccard similarity is close to 0 as the approximate calculation shows: $\frac{P(a_{N}a_{i})}{P(a_{N}) + P(a_{i}) - P(a_{N}a_{i})} \approx \frac{P(a_{N}a_{i})}{P(a_{N})} \approx 0$. Based on this observation, we filter out the alert pairs that have a small Jaccard similarity to perform denoising. 

\vspace{-10pt}

\subsubsection{Leveraging Spatial Relations}

To measure the correlation of alerts in service topology, we introduce the historical links between services for correlation mining. The owning service is one of the attributes of alerts, which can be obtained once the alert is triggered. With the services and historical failures, we can construct the topology graph of services. The nodes in this graph are services, and edges indicate that there are correlated alert pairs from these two services in historical records. The edge points from the service of the earlier alert to the service of the later alert. Besides, each service may trigger multiple alerts in one failure, which means there are multiple alerts in each node in the graph. This also requires a different strategy from the existing graph embedding methods when the graph is sampled.

Inspired by node2vec~\cite{grover2016node2vec}, a novel node representation method for graphs, we construct a two-stage service topology pattern mining approach. We first sample the topology graph by an alert-aware random walk technique, then utilize the skip-gram model for alert embedding learning. Finally, the embedding of alerts are used for similarity computing with the distance of alert embeddings. 

In the sampling stage, we apply the random walk on the service topology graph. The sampled results are node sequences in this approach. However, there are multiple alerts in one single node. Thus we randomly sample one of those alert to the sequence. 
The sampling direction can be controlled in this algorithm, reaching a balance between exploring different areas like breadth-first search (BFS) and focusing on local neighborhoods as depth-first search (DFS). In our scenario, we observe the failure can be propagated to services of other components, which are far from the original node in the topology graph. To extract the long propagation characteristics, we set up the sampling direction similar to DFS.

In the embedding stage, we employ the skip-gram model~\cite{mikolov2013distributed}. The skip-gram model is first designed in natural language processing tasks, to predict the missing word given the neighboring context. It maximizes the likelihood of predicting the context nodes using the embedding of targets. In our task, we focus on the embedding of alerts in the sequence. The embedding of alerts contains the topology features of their positions and neighbors in the graph, which is useful for correlation mining. Therefore, the embeddings are representative of the spatial relation of alerts that can be used for similarity computing. For the alert pairs $a_{1}$ and $a_{2}$, we obtain their embedding from the skip-gram model. If the alert is unseen to the model, we compute the average embedding of the owning service instead. Then we obtain the distance of these embeddings as a similarity score for correlation denoted as $S_{a_{1}a_{2}}$.  

\subsubsection{Combining Temporal and Spatial features}
We obtain the temporal correlation metric $T$ by conditional possibility, and get the spatial correlation metric $S$ by the distance of embeddings. To combine the two metrics, we compute the final similarity score by a linear combination:
\begin{align*}
similarity\ score = max\{T_{a_{1}|a_{2}},\ T_{a_{2}|a_{1}}\} - \alpha \cdot \frac{S_{a_{1}a_{2}} - S_{min}}{S_{max} - S_{min}}
\end{align*}
The temporal metric $T$ is a possibility value ranging from 0 to 1, while the spatial metric is a positive real number that stands for distance. To align with the temporal metric, we normalized the spatial metric to the same interval. Notice that the larger distance represents a lower similarity, thus the spatial term is negative. The parameter $\alpha$ controls the threshold for correlation. We conduct a grid search, with $\alpha$ ranging from 0 to 10 and the step size as 0.5, and determine the optimal $\alpha$ is 3.5 in our task. Finally, if the similarity score is positive, we output the alert pairs as correlated. Otherwise, we input the pair to the next module.

\subsection{LLM Reasoning Module}

In \stnm, we output the alert pairs with positive similarity scores. For those pairs where the similarity score is negative, they can be irrelevant, or they are not statistically similar but logically correlated. Statistical methods cannot get better conclusions on their relevance. Thus we need to introduce more knowledge for alerts, and utilize more advanced but relatively inefficient techniques (\eg LLM) for further analysis.

As section~\ref{sec:background} mentioned, SOP documents are widely used for service failure handling. SOP contains more key information about alerts, including possible causes, explanations, impacts on the system and recommended mitigation steps. Furthermore, those contents contain potential domain knowledge. For example, the impacts and mitigation steps usually describe the fact that hardware outage may be the root cause of customer-side service failure. However, the SOPs are written in unstructured natural language with long lengths. It is hard to extract key information from SOP automatically. With the capability of context understanding and reasoning, LLMs become one of the feasible solutions to SOP processing. We design a two-round interaction with LLM based on the COT technique. In the first round, we summarize and store the key information. Then in the second round, we design the prompts for the in-context learning (ICL) method with the matched historical samples and pre-defined rules. 
And we further fine-tune the LLM by the p-tuning v2 ~\cite{liu2021p} technique, a popular supervised fine-tuning (SFT) method.
With the extra model parameters trained in p-tuning v2,  we can get responses to the alert correlations accurately and clearly.

\subsubsection{Knowledge Extraction}
To understand and leverage the domain knowledge in the SOPs, we introduce LLMs for semantic information extraction. LLMs are capable of natural language understanding, but the long text input would worsen the effectiveness of context handling. Most of the existing LLMs limit the length of the prompt, which is not enough for the contents and SOPs of the alert pair along with the query instruction. 
To maintain the key information of documents and shorten the length of the input, inspired by the COT technique, we interact with LLMs by rounds, where the result of the current round would be inserted as input for the next round. This helps LLM focus on the task of the current round and preserves the information in previous rounds.

\begin{figure}[t]
  \centering
  \includegraphics[width=0.45\textwidth]{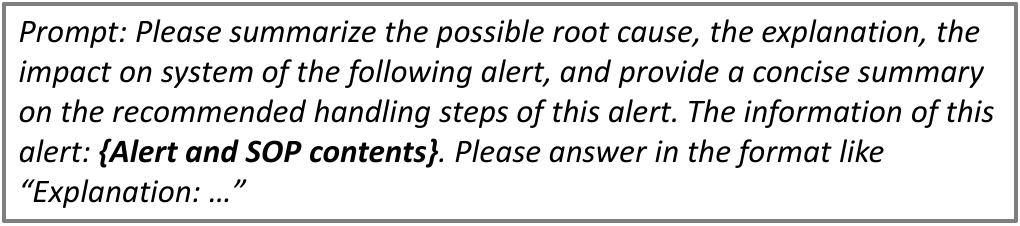}
  \vspace{-5pt}
  \caption{The prompt template for knowledge extraction}
  \label{fig:prom_extr}
  \vspace{-15pt}
\end{figure}

In particular, we perform knowledge extraction on the original SOPs, summarize the main knowledge of these documents, and store the output for the next round interaction as the upper row of Figure~\ref{fig:fr} shows. We design the prompt for this step by mainly asking LLM to summarize the information in the given SOPs, and stress out some aspects that are useful for reasoning, such as the detailed explanation, the consequence of the system and the handling steps. The answers follow the instructions, containing the attributes given in the prompt as in Figure~\ref{fig:prom_extr}. 

\subsubsection{In-Context Learning}
With the extracted SOP information and domain knowledge obtained in the knowledge extraction, several approaches are proposed for LLM to take up the knowledge. In-context learning (ICL) is one of the most popular prompting methods. As a few-shot learning method, ICL requires a few preprocessed samples, which is flexible on deployment in real-world scenarios. Recent research~\cite{gao2023what} shows that the ICL method with high-quality samples can achieve a similar performance compared with tuning methods. Therefore, we adopt ICL as one of the methods to enhance LLM reasoning.

We construct the prompts for ICL with the three parts of samples, query and rules as the LLM reasoning component in Figure~\ref{fig:fr} shows.
Samples of semantically similar previous alerts and SOPs provide a reliable reference for the current handling.
Thus the quality of the samples has a significant impact on the response from LLM. 
To extract the semantic features and compute the similarity directly, we employ the FastText~\cite{joulin2016bag} as our embedding model. FastText is a lightweight model that performs well on domain-specific texts, but is more efficient in training and evaluating compared with the generalized embedding model such as BERT~\cite{devlin2018bert}. We embed the alert title and SOP documents into a 750-dimension vector. For each alert pair in \smnm, \smnm embeds the two alerts, respectively. Then compute the semantic similarity with the vector distance of the current alert and all sampled alerts, taking the \textit{top-k} similar samples to be inserted into the prompt. Due to the input length limit of LLM, we adopt the top-1 similar positive sample and top-1 negative sample as final samples in prompts.

Query organizes the main question of the prompt. It contains the title and summarized SOP of two alerts, and connects the information fluently with natural language. The query part also consists of some instructions to highlight the core problem, suggesting the structure and format of the answer. Thus, the responses from LLM would be in a more standard format which is efficient to evaluate.

\begin{figure}[t]
  \centering
  \includegraphics[width=0.4\textwidth]{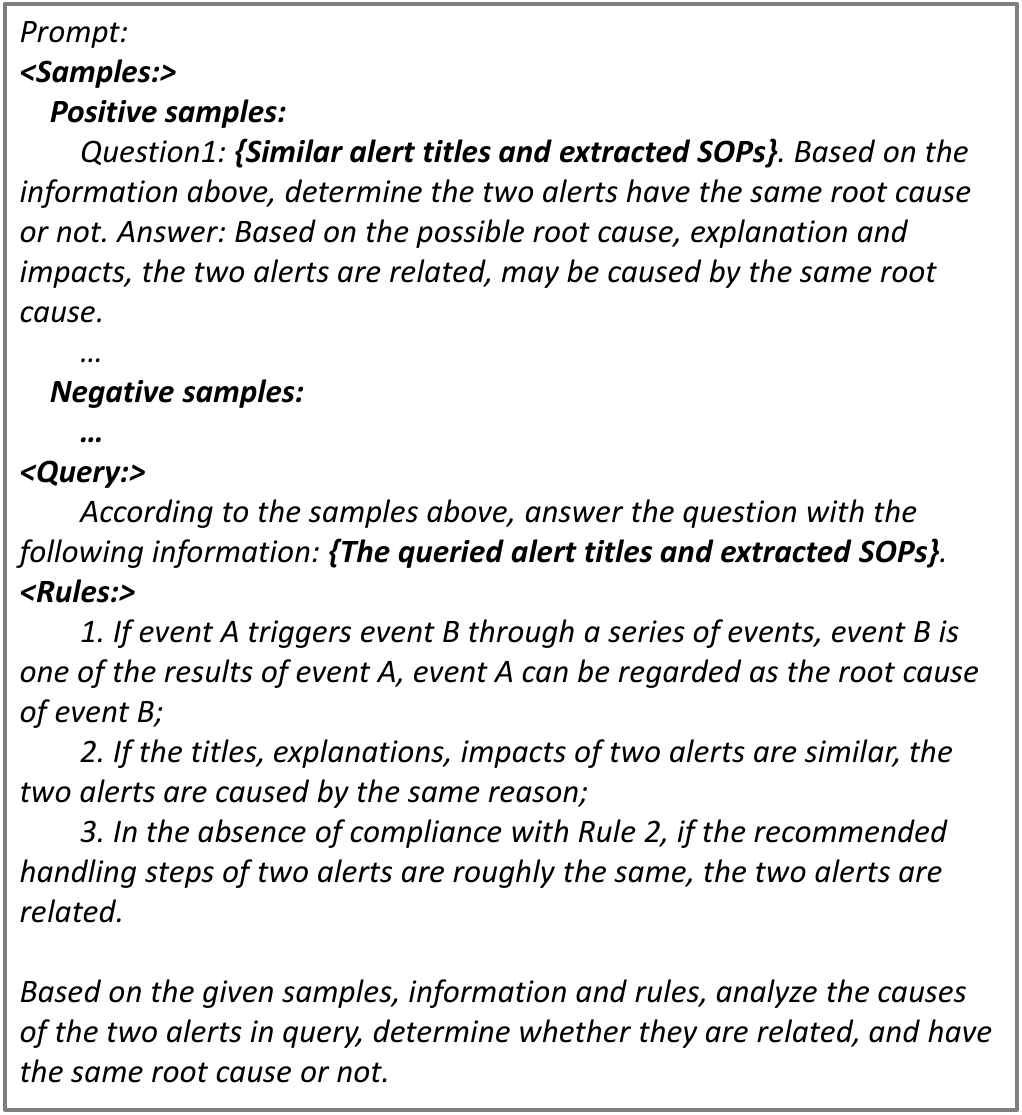}
  \caption{The prompt template for ICL}
  \label{fig:prom_icl}
\end{figure}

Rules help LLM generate more stable responses.
The general knowledge implied in LLM makes it respond flexibly to some concepts like "\textit{root} cause" and "\textit{similar}", resulting in unstable answers to the same question. To improve the consistency of the responses, we insert some pre-defined rules for reasoning as Figure~\ref{fig:prom_icl} shows. Rule 1 explains the propagation feature of causality to help LLM understand the concept of root cause. Rule 2 and 3 explicitly define the priority of different information when comparing similarity.

\subsubsection{Supervised Fine-Tuning}
Another method for LLM to learn domain knowledge is to retrain the parameters in the model. However, it is impractical to fully update the billions of parameters in LLMs due to resource constraints. Thus fine-tuning techniques are proposed for updating a small part of the parameters, preserving the ability of natural language processing and obtaining domain-specific knowledge. P-tuning~\cite{liu2021p} is one of the most popular methods for LLM fine-tuning. Different from updating parameters from LLM, p-tuning maintains another set of parameters as the lower row in Figure~\ref{fig:fr} shows, to transform prompts into embedding. By learning these extra parameters, p-tuning makes LLM respond more consistently with prompts, and get knowledge from the training dataset. In \smnm, we employ p-tuning v2~\cite{liu2022p}, which provides a larger scale of parameters with new learning skills, leading to a more efficient and effective tuning process.

To perform fine-tuning on LLM, we split part of the data from the training set to make up the validation set. The training set covers alerts and SOPs reported from January to May of 2023, consisting of about 85.1\% of all data. We randomly select data from the original training set to the amount of 80\% as the new training set, and the rest 5.1\% is for the validation set. We perform p-turning on GPUs and train the model for 1800 steps. We evaluate the fine-tuned model on the validation set and save every 300 steps. Finally, the checkpoint with minimum validation loss is chosen for testing. 

\section{Evaluation}\label{sec:evaluation}
\noindent We evaluate \frnm by answering the research questions (RQs).
\begin{itemize}[leftmargin=*, topsep=0pt]
    \item RQ1: What is the effectiveness of \frnm ?
    \item RQ2: What is the impact of each component in \frnm ?
    \item RQ3: What is the time efficiency of \frnm ?
\end{itemize}

\subsection{Datasets}

The three datasets in our evaluation are collected from the industrial production environment of \huawei. The datasets contain alerts from 2023/01/01 to 2023/06/30, covering over 60 services from 14 physical regions. There are around 500,000 alerts and 3,000 identical SOP documents in all three datasets. Alerts and SOPs from 2023/01/01 to 2023/05/31 are used to select parameters in correlation mining, ICL samples supporting and LLM fine-tuning. The remaining data are used for the testing of our framework.

\begin{table*}[]
\small
\captionsetup{justification=centering}
\centering
\caption{Effectiveness of experimental approaches on identifying correlated alerts}
\vspace{-12pt}
\label{tab:effectiveness}
\begin{tabular}{cccccccccc}
\toprule
\multirow{2}{*}{Method} & \multicolumn{3}{c}{Dataset A} & \multicolumn{3}{c}{Dataset B} & \multicolumn{3}{c}{Dataset C} \\
 & Precision & Recall & \textbf{F1 score} & Precision & Recall & \textbf{F1 score} & Precision & Recall & \textbf{F1 score} \\ \midrule
FP-Growth & 0.413 & 0.744 & 0.531 & 0.409 & 0.692 & 0.514 & 0.364 & 0.655 & 0.467 \\
DBSCAN & 0.166 & 0.432 & 0.240 & 0.187 & 0.364 & 0.247 & 0.227 & 0.400 & 0.289 \\ 
\midrule
Alert Storm & 0.328 & 0.662 & 0.438 & 0.356 & 0.604 & 0.448 & 0.338 & 0.639 & 0.442 \\
LiDAR & 0.636 & 0.514 & 0.568 & 0.682 & 0.589 & 0.632 & 0.672 & 0.653 & \uline{0.662} \\
OAS & 0.454 & 0.483 & 0.468 & 0.427 & 0.521 & 0.469 & 0.486 & 0.561 & 0.521 \\
iPACK & 0.691 & 0.635 & 0.661 & 0.603 & 0.642 & 0.621 & 0.654 & 0.614 & 0.633 \\ \midrule
\frnm w/o SFT & 0.694 & 0.651 & \uline{0.672} & 0.638 & 0.693 & \uline{0.664} & 0.653 & 0.638 & 0.645\\
\frnm & 0.892 & 0.924 & \textbf{0.908} & 0.916 & 0.943 & \textbf{0.930} & 0.921 & 0.882 & \textbf{0.901}\\
\bottomrule
\vspace{-15pt}
\end{tabular}
\end{table*}

\subsection{Baselines}
We compare our approaches with two traditional machine-learning techniques and four popular recent research methods. FP-growth~\cite{han2000fpgrowth} is an algorithm used for frequent pattern mining in data mining and association rule learning. DBSCAN (Density-Based Spatial Clustering of Applications with Noise)~\cite{ester1996dbscan} is a popular density-based clustering algorithm that can discover semantic clusters with alert title embeddings. The four state-of-the-art research methods are introduced as follows:

\textbf{AlertStorm~\cite{zhao2020alertstorm}:} AlertStorm conducts the first empirical study of the alert storm and proposes a novel approach to handling alert storms, consisting of alert storm detection and alert storm summary. The detecting part adaptively and accurately detects alert storms based on Extreme Value Theory, while the summarizing part proposes an alert denoising method and an alert discrimination method. We mainly focus on its summarizing part in our evaluation.

\textbf{LiDAR~\cite{chen2020lidar}:} LiDAR (Linked Incident identification with DAta-driven Representation) consists of a textual embedding module for semantic information and a component representation learning module for dependency structure. It is an integrated framework that can effectively identify possible linked incidents. In our evaluation, we managed to utilize LiDAR for identifying linked alerts by providing similar semantic and dependency information.

\textbf{OAS~\cite{chen2022oas}:} OAS (Online Alert Summarizing) first aggregates the contextual information of alert words by the word frequency in alert contents, then mines the common behavior pattern between alerts from the alert occurrence series. Finally, it combines the above two types of alert information and determines the correlation by a deep learning model.

\textbf{iPACK~\cite{liu2023ipack}:} iPACK is an incident-aware method for aggregating duplicate tickets, consisting of alert parsing, incident profiling and ticket-event correlation. In particular, incident profiling filters noisy events and links the correlated events that are caused by the same incidents. In our evaluation, we apply incident profiling on alerts to determine correlated alerts.

\subsection{Implementation Details}
We implement \frnm on a Windows server with an Intel(R) Core(TM) i7-10700 CPU @ 2.90GHz and 32GB RAM. The inference tasks and fine-tuning of LLM components are done on 4 NVIDIA Tesla T4 GPUs with 12GB graphic memory for each. We set the parameters of fine-tuning as default $PRE\_SEQ\_LEN = 128$ and $LR = 2e-2$. For the compared methods which are not open-sourced, we follow the demonstrated techniques and given parameters to reproduce the works and ensure their performance as well.

\subsection{Metrics}
\label{subsec:metric}
The goal of our methods is to identify the correlated alerts. Some of the compared methods score on the alert pairs to determine the result, while the others directly give out the aggregated alert groups by clustering methods. Thus a unified output format is required. For those clusters, we transform them into alert pairs by pair-wised combination. In particular, for a $m$-alert cluster $Cluster=\{a_{i}\}, i \in [1,2,3...,m]$, we obtain $m^{2}-m$ alert pairs as $Pairs = \{(a_{p}, a_{q})\}, p, q \in [1,2,3...,m],\ p\neq q$.
To compare the pair-wised results with the ground-truth labels, we compute the four metrics of the fusing matrix and obtain the final measurement: Ture Positive (TP) stands for the correlated alerts in results which are also labeled as related in ground truth; True Negative (TN) are for those irrelevant alert pairs both in results and ground truth. False Positive (FP) are correlated alerts in our tests but are not correlated in ground truth. False Negative (FN) means the irrelevant pairs in tests but actually correlated in ground truth.
With these four basic metrics, we obtain the precision, recall and F1-score as the final metrics for our evaluation, where $precision = \frac{TP}{TP + FP}$, $recall = \frac{TP}{TP + FN}$, $F1-score = 2\cdot\frac{precision\cdot recall}{precision + recall}$.

\subsection{Evaluation Results}
\subsubsection{\textbf{RQ1: }}
In this RQ, we demonstrate the effectiveness of \frnm on identifying correlated alerts by comparing it with the two machine learning algorithms and four research methods. The evaluation result is shown in Table~\ref{tab:effectiveness}. The highest F1-score is shown in \textbf{bold}, which represents the best overall performance considering the correctness and completeness of the result. And the second-highest F1-score is \uline{underlined}. 

We implement the LLM module of \frnm with ICL and SFT independently. We evaluate the two methods and list the result in Table~\ref{tab:effectiveness}. \frnm without SFT reaches the best F1-score in Dataset A and B, while getting the second highest in Dataset C, compared with the existing methods. The result supports that LLM contributes to the correlation analysis. However, the increase of F1-score is not significant, which is 1.7\% and 5.1\% compared to the state-of-the-art method on Dataset A and B, respectively. This result shows the great potential to improve ICL on the alert aggregation task. There are several possible reasons for the low improvement: 1) The parameters of the internal LLM are insufficient, compared to the commercial LLMs with over 100B parameters. 2) The given samples in the prompt are not enough, due to the length limits of LLM inputs. 3) The quality of samples is not satisfying, because the sample matching is based on a simple SOP embedding. 4) The sample response only shows the label, without giving any reasoning steps which makes LLM hard to follow.

\frnm achieves the best F1-score among all datasets, and outperforms the existing method by 37.3\%, 47.1\%, 36.1\% on Dataset A, B, C, respectively. The distinguished improvement in the F1-score indicates the p-tuning parameters have learned the domain knowledge and can be effectively utilized by LLM for inference. However, the training set size is relatively small because the labeled correlated alerts and SOPs are limited. Thus there is a potential overfitting risk during the fine-tuning, and the learning steps and validation loss should be carefully checked.

As for the baseline methods, the recall of FP-Growth, DBSCAN and Alert Storm are significantly higher than their precision. Because those three methods output their result by clusters instead of the identified correlated pairs, and we transform the clusters into pairs for our evaluation. Thus the number of false-positive samples would be enlarged, resulting in a low accuracy. Both LiDAR and OAS mine the correlation through semantic information and co-occurrence information. However, the pattern mining part of OAS is not so effective in our evaluation, because our data do not contain the \textit{Type} or \textit{Template} attributes. Thus the performance of OAS is lower than expected. The relation learning module in iPACK utilizes point-wise mutual information (PMI) as the metric, which is mathematically similar to the conditional possibility in our design. Besides, the Kneedle algorithm of iPACK for denoising is effective for the evenly distributed noises in our task. Thus iPACK reaches a close performance with LiDAR and the \frnm without SFT.

\begin{table*}[]
\small
\captionsetup{justification=centering}
\centering
\caption{Effectiveness of each component in \frnm on identifying correlated alerts}
\vspace{-12pt}
\label{tab:ablation}
\begin{tabular}{cccccccccc}
\toprule
\multirow{2}{*}{Method} & \multicolumn{3}{c}{Dataset A} & \multicolumn{3}{c}{Dataset B} & \multicolumn{3}{c}{Dataset C} \\
 & Precision & Recall & \textbf{F1 score} & Precision & Recall & \textbf{F1 score} & Precision & Recall & \textbf{F1 score} \\ \midrule
w/o temporal relation & 0.521 & 0.583 & 0.550 & 0.536 & 0.566 & 0.551 & 0.548 & 0.579 & 0.563\\
w/o spatial relation & 0.876 & 0.893 & 0.884 & 0.861 & 0.852 & 0.856 & 0.874 & 0.853 & 0.863\\ 
w/o LLM & 0.652 & 0.596 & 0.616 & 0.594 & 0.633 & 0.613 & 0.627 & 0.649 & 0.638 \\
\frnm & 0.892 & 0.924 & 0.908 & 0.916 & 0.943 & 0.930 & 0.921 & 0.882 & 0.901 \\
\bottomrule
\vspace{-18pt}
\end{tabular}
\end{table*}

\subsubsection{\textbf{RQ2: }}
In this RQ, we illustrate the effectiveness of each component in \frnm by removing the corresponding parts from the framework and performing the evaluation. The full version of \frnm combines the correlation mining module, which leverages temporal and spatial relations, with the LLM reasoning module utilizing the SFT technique. To remove the contribution of temporal relation, we only maintain the sequence sampling and node embedding for spatial relation in the correlation mining module, while the LLM module remains unchanged. And vice versa for the spatial relation. To remove the contribution of LLM, we mark all the alert pairs with negative similarity scores as unrelated to get the test result.

The evaluation result is shown in Table~\ref{tab:ablation}. The average F1-score reduction on all datasets for removing temporal relation, spatial relation and LLM module is 39.3\%, 5.5\%, 31.8\%, respectively. 
The contribution of spatial relation is much smaller than the other two components. We conclude two main reasons from our historical data for the low effectiveness: 1) Inaccuracy: We construct the service topology graph by linking the related service nodes in history. However, the different intensity of service connection is not taken into consideration. 
Some service pairs appear more frequently than other pairs, which indicates a closer correlation and yields a higher contribution to spatial similarity. 
2) Incompleteness: The historical data can not cover the real-world topology, and can not deal with the new services and alerts as the system updates from time to time. We expect the performance of spatial relation would be improved with high-quality topology data.

The temporal relation and LLM module account for the main performance of \frnm. For the temporal component, the distribution of alerts in the online service system follows the temporal locality. Thus the time widow division method is suitable and effective for capturing the locality. Besides, the correlation analysis based on co-occurrence pattern mining is reliable and achieves satisfying performance in previous research on many tasks~\cite{wang2021fast, wang2021groot, li2021fighting}.
The LLM component utilizes the semantic information to determine correlation. Different from previous semantic methods based on embeddings, LLM learns the domain knowledge from SOPs, and stores it in local parameters. The domain knowledge helps LLM understand alerts better and makes LLM possible to infer and compare the cause of alerts. In addition, LLM acts as the complementary of co-occurrence mining, further improving the performance by analyzing the statistically unrelated samples. For example, some correlated alerts may have a large time interval between their appearances, and some may be triggered rarely or are completely new and unseen to the model. LLM tackles these samples by summarizing their SOPs and analyzing the causes, thus achieving significant improvement over statistical methods. 

\begin{table}[]
\centering
\small
\caption{Average inference time (s)}
\vspace{-6pt}
\label{tab:efficiency}
\begin{tabular}{ccccc}
\toprule
\textbf{Method} & \textbf{Dataset A} & \textbf{Dataset B} & \textbf{Dataset C} \\ \midrule
Alert Storm & 2.14 & 1.67 & 1.64  \\
LiDAR & 1.12 & 0.78 & 0.82\\
OAS  & 0.19 & 0.29 & 0.23\\
iPACK & 0.47 & 0.52 & 0.58\\
\midrule
ICL & 42.81 & 50.02 & 47.86 \\
\frnm & 5.78 & 8.94 & 7.48\\
\bottomrule
\end{tabular}
\vspace{-15pt}
\end{table}

\subsubsection{\textbf{RQ3: }}
In this RQ, we investigate the efficiency of \frnm from the time cost of offline training and online inference. For the offline training stage, the different components of \frnm can be trained in a paralleled way. The most time-consuming component is the LLM. The SFT takes hours to fine-tune depending on the parameter settings and hardware environments. In practice, the time cost of training would not reduce the efficiency. Because the fine-tuned model only needs to be trained once before being deployed.
The time cost of fine-tuning is about 12 hours in our evaluation. That means if it is necessary to retrain parameters due to system updates after a long time, the time cost is practical and acceptable. The retraining process would not interrupt the on-work online inference. 

For the time cost of online inference, the average inference time for each alert pair is shown in Table~\ref{tab:efficiency}. The ICL is the most time-consuming technique and much more costly compared with other methods, which matches the observation of the low-efficiency problem of LLM. 
Even worse, the long ICL prompt with samples, queries and rules results in a longer time to wait for the LLM response.
AlertStorm and LiDAR give a result with 1.82 and 0.91 seconds on average. The main cost for Alert Storm is the DBSCAN clustering, and for LiDAR is the two types of embedding computing with distance calculating. OAS and iPACK response in less than 1 second for all datasets. In our implementation, the co-occurrence pattern mining part in OAS is not fully implemented because there is no attribute for the alert \textit{type} or \textit{template} in our dataset. The main technique we reproduce from iPACK, incident profiling, is a statistic method with high efficiency. Thus the time costs of these two methods are low. 
As for the \frnm, the average response time is 7.4 seconds. 
The LLM module in \frnm accounts for the most time cost. However, \frnm significantly reduces the time cost compared with the standalone ICL method. Because many of the alerts are filtered and denoised by the \stnm , which is an efficient statistic component. LLM only handles the rest of the alerts. Thus even with the LLM introduced in \frnm, the inference time cost is still at a similar scale to the state-of-the-art designs which contain both statistic and semantic modules like LiDAR. 

\section{Industrial Experience}
\label{sec:indst}

In this section, we share our experience in integrating \frnm to our SRE (site reliability engineering) platform in \huawei.
The SRE platform is a centralized platform that manages the life cycle of alerts, including alert configuration, alert reporting, assignment, mitigation and resolution.
One critical object for OCEs using the platform is to ensure a cloud problem can be quickly resolved to minimize customer impact.
However, this is usually challenging when OCEs face overwhelming information on the platform. 
Especially, when an alert storm happens, hundreds of alerts can be simultaneously reported from multiple services.
In particular, during alert storms, numerous alerts can be simultaneously reported from multiple services. 
OCEs often find it necessary to aggregate alerts stemming from the same problem to accelerate their resolution. 
In our previous practice, multiple OCEs were required to manually comprehend these alerts and identify correlated instances through discussions. 
This approach proved to be inefficient, labor-intensive, and heavily reliant on the OCEs' experience.
Considering the large scale of \huawei, it is non-trivial for OCEs to understand these alerts and correlate them quickly.
This is primarily due to each OCE primarily specializing in a specific component of a service, thereby limiting their knowledge scope.

\begin{figure}[t]
  \centering
  \includegraphics[width=0.4\textwidth]{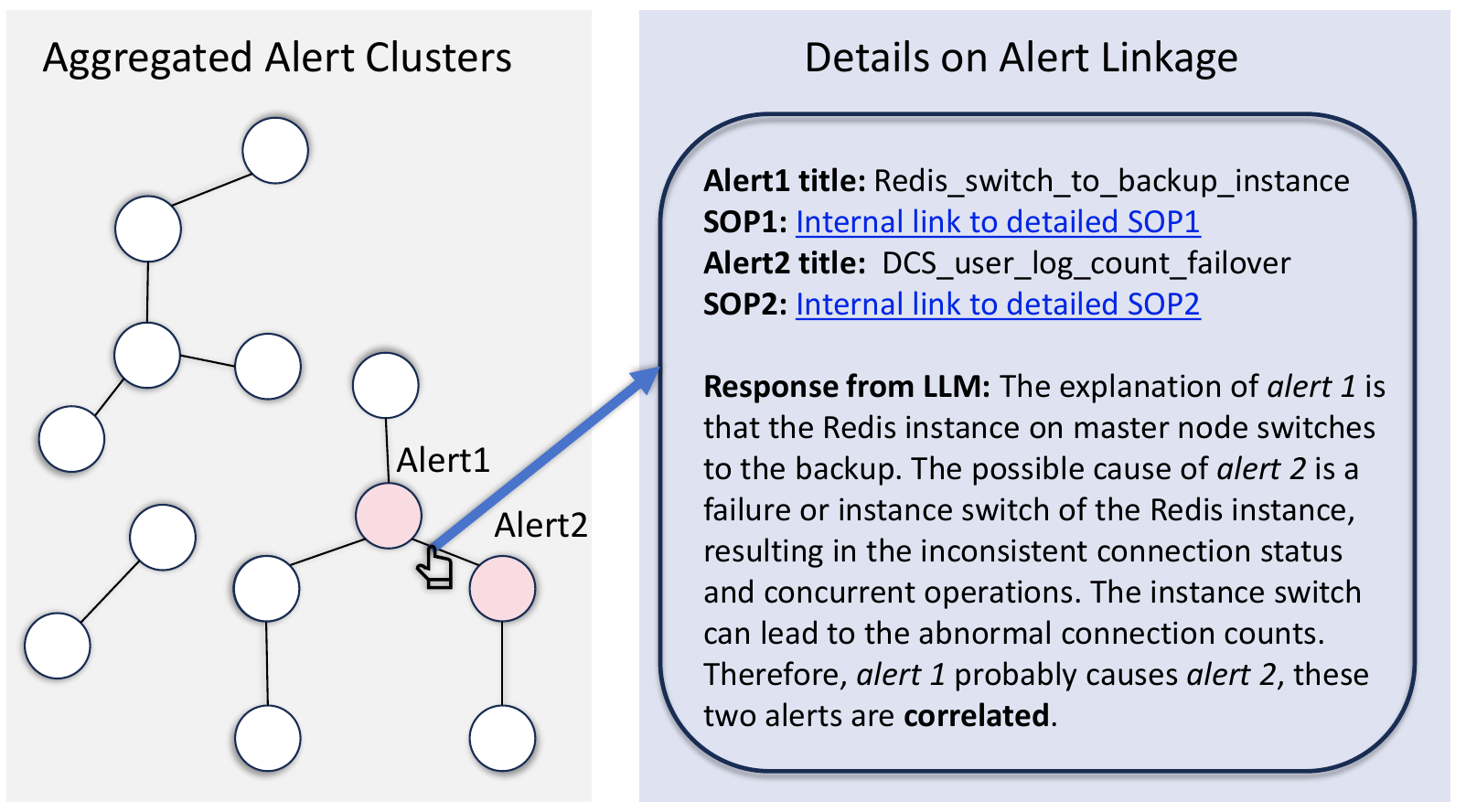}
  \vspace{-10pt}
  \caption{An industrial case of alert clusters}
  \label{fig:indst_exp}
  \vspace{-20pt}
\end{figure}

To assist OCEs in alert analysis, we have integrated \frnm within the SRE platform. \frnm has learned from extensive knowledge (\eg SOPs) accumulated in the SRE platform, which covers almost all services in \huawei. 
As demonstrated in Figure~\ref{fig:indst_exp}, \frnm will present alerts in clusters to OCEs, where the link between alerts is associated with linkage details \ie the concrete title, associated SOPs and response from LLM. 
Alerts that are not correlated do not possess such linkages.
In practice, \frnm not only is capable of predicting the relationships between two alerts for aggregation, but it can also provide interpretable explanations based on the knowledge it has acquired, such as summarized information from the SOPs and its reasoning procedure.
In \huawei, we have received positive feedback from OCEs. With the aggregation results, OCEs can quickly obtain a comprehensive picture of a failure event and improve the efficiency of resolving a problem. 
More importantly, OCEs found it is more acceptable to interpretable results rather than yes or no predictions. 
They appreciate that the adoption of \frnm has effectively bridged knowledge gaps that exist across different services, contributing to the overall efficiency of cloud maintenance activities.

\section{Threats to Validity}
We identify the following threats to external and internal validity.

\noindent\textbf{External validity.} 
The study’s object is the primary external threat. In the evaluation, we utilize the Standard Operation Procedure (SOP) documents. The three datasets are all from \huawei, because there is no publicly available dataset containing such detailed and complete alert descriptions as SOPs. However, \huawei is a globally recognized cloud provider, known for its extensive scale and leadership in the industry. And the data are indiscriminately collected from various services of diverse regions. Thus the evaluation should be representative and convincing. Furthermore, the historical data are maintained and accumulated regularly in commercial cloud providers. The contents of SOPs are about necessary alert handling, which is commonly recorded in related documents such as failure reports of various cloud systems. Thus we believe \frnm is capable of generalizing to similar cloud systems.

\noindent\textbf{Internal validity.} 
Implementation and parameter settings are the primary internal threats. For implementation, the compared research methods are not open-sourced. Thus we reproduce the methods based on the original paper, following the detailed algorithms and parameter settings. We utilized established libraries to implement the core components. Peer code review is employed for the proposed and baseline methods implementations. For parameter settings, we conduct a thorough hyperparameter tuning process, such as grid-search, to select the optimal results. 

\section{Related Work}

\subsection{Failure Analysis}
Failure analysis, such as incident management and root cause analysis, has become a popular topic as online service systems develop rapidly nowadays~\cite{chen2020towards, chen2019empirical, chen2020incidental}. It aims to mitigate the failure timely to recover the service quality, and identify the root causes to diagnose and fix the outage thoroughly. Researchers have devoted sustained efforts to reliable analysis based on various data sources, such as alerts, incidents and internal failure reports. For alerts, Zhao \etal \cite{zhao2020alertstorm} observe and name the \textit{alert storm} phenomenon. They propose a learning-based method to denoise the alerts by the isolation forest, and design a clustering-based method to group the alerts by DBSCAN. OAS~\cite{chen2022oas} combines the semantic and behavior information, and fuse the representation of two types of information by a deep learning model. iPACK~\cite{liu2023ipack} mines the correlation pattern of parsed alerts by point-wise mutual information. These methods mainly focus on textual information and co-occurrence patterns in time windows, while the service topology is not well utilized. For incidents, LiDAR~\cite{chen2020lidar} proposes a textual encoding module with TextCNN~\cite{kim2014convolutional} for incident contents, a component embedding module with Skip-Gram~\cite{mikolov2013distributed} for service topology, and combine them to identify the linked incidents. 
COT~\cite{wang2021fast} predicts the root cause by building service correlation from incident correlation.
Groot~\cite{wang2021groot} analyzes the root cause by constructing dependency and causality graphs based on failure events. Warden~\cite{li2021fighting} groups the incidents with the same cause to perform timely incident triage. 
For internal failure reports, Shetty \etal ~\cite{shetty2022autotsg} from \textit{Microsoft} leverage the \textit{Troubleshooting Guides} (TSG) to propose AutoTSG, for automatically minimizing the manual effort of On-call engineers on incident handling. Different from these works, we introduce the internal \textit{Standard Operation Procedure} (SOP) documents in \huawei and further leverage the Large Language Model (LLM) to perform reasoning based on the alert and SOP contents.

\subsection{LLM for Software Engineering}
With the rapid development and promising rise of LLM in recent years, LLM has been employed to solve various challenging problems in software engineering, including code generation, summarization, testing and root cause analysis. Mastropaolo \etal \cite{mastropaolo2021studying} perform a study about the capability of fine-tuned text-to-text-transfer-transformer (T5) on a series of code-related tasks. LANCE~\cite{mastropaolo2022using} generates logging statements by fine-tuning the T5 model. On online systems, Ahmed \etal \cite{ahmed2023recommending} do the first large-scale study to evaluate the effectiveness of recommending root cause and mitigation steps by fine-tuned \textit{GPT-3.x}. RCACopilot~\cite{chen2023empowering} collects the runtime diagnostic information, then predicts the root cause category of incidents with the collected information by LLM. OASIS~\cite{jin2023assess} aggregates the incidents and generates human-readable summaries by fine-tuned \textit{GPT-3.x}. Different from those methods, we introduce the fine-tuned internal LLM to perform alert aggregation based on domain knowledge reasoning.

\section{Conclusion}
 In this paper, we present \frnm, a novel hybrid approach for online alert aggregation in large-scale cloud systems. By leveraging detailed knowledge from Standard Operating Procedures (SOPs), COLA effectively aggregates alerts that may be semantically dissimilar or lack sufficient statistics for analysis.
 The proposed framework combines a correlation mining component and an LLM reasoning component, allowing for efficient handling of a large volume of alerts in practical scenarios.
 Extensive evaluation on three real-world datasets from \huawei demonstrates that COLA outperforms state-of-the-art methods in terms of F1-scores while maintaining comparable efficiency. 
 The deployment of COLA in the production environment of \huawei further validates its effectiveness. 
\section{Acknowledgment}
The work described in this paper was supported by the Research Grants Council of the Hong Kong Special Administrative Region, China (No. CUHK 14206921 of the General Research Fund). 

\balance
\bibliographystyle{ACM-Reference-Format}
\bibliography{ICSE24}
\end{document}